\pdfoutput=1
\documentclass[]{aa}
\usepackage{amsmath}
\usepackage{latexsym}
\usepackage{longtable}
\usepackage{natbib}
\usepackage[latin1]{inputenc}
\usepackage{graphics}
\usepackage{epsfig}
\usepackage{graphicx}

\begin{document}

\title{The power spectrum of the cosmic microwave background Kolmogorov 
maps: possible clue to correlation of voids}
\author{V.G.Gurzadyan\inst{1}, 
A.L.Kashin\inst{1}, H.G.Khachatryan\inst{1}, A.A.Kocharyan\inst{1,2}, E.Poghosian\inst{1}, D.Vetrugno\inst{3},
 G.Yegorian\inst{1}}

\institute
{\inst{1} Yerevan Physics Institute and Yerevan State University, Yerevan,
Armenia\\
\inst{2} School of Mathematical Sciences, Monash University, Clayton, Australia\\
\inst{3} University of Lecce, Lecce, Italy
}

\date{Received (\today)}

\titlerunning{CMB}

\authorrunning{V.G.Gurzadyan et al}

\abstract{
The power spectrum is obtained for the Kolmogorov stochasticity parameter map for 
WMAP's cosmic microwave background (CMB) radiation temperature datasets. The interest for CMB Kolmogorov map is that
it can carry direct information about voids in the matter distribution, so that the correlations
in the distribution of voids have to be reflected in the power spectrum. 
Although limited by the angular resolution of the WMAP, this analysis shows the possibility of acquiring this crucial information via CMB maps. Even the already obtained behavior, some of which is absent in the simulated maps, can influence
the development of views on the void correlations at the large-scale web formation.
}


\maketitle

\section{Introduction}
Correlations in the full sky or large enough surveys contain clues to the early Universe and its present structure. The acoustic peaks of the cosmic microwave background (CMB) power spectrum revealed a set of cosmological parameters with particular accuracy \cite{dB,WMAP5}. The baryon acoustic oscillations (see Percival et al 2009) are crucial for studies of the formation of the large-scale structure, including the role of dark matter and dark energy.    

Below, we construct the power spectrum of a novel type of full sky map, those representing the distribution of the Kolmogorov stochasticity parameter of the CMB temperature maps. Kolmogorov's parameter is a descriptor for a degree of randomness \cite{Kolm,Arnold} and when applied to the CMB temperature datasets results in a map (K-map) \cite{GK2009} that has both features resembling the temperature maps, like the outlined Galactic disk, but also ones with different contents. The Cold Spot \cite{Cruz}, the non-Gaussian structure of negative mean temperature, was noticed thanks to the excess of the K-parameter with respect to its mean value over the sky. Moreover, the behavior of the K-parameter, i.e. of the degree of the randomness was increasing towards the boundary of the Cold Spot (Gurzadyan and Kocharyan 2008, 2009). Both features are compatible to the void nature of the Cold Spot. Other spots and regions have been noticed in the K-map, which are studied with other descriptors as well (Rossmanith et al 2009), and other noticed non-Gausianities can also be among the applications (Gurzadyan et al 2005, 2008).  

If the Kolmogorov CMB map is able to reflect the features in the matter distribution, it is therefore natural to study the large-scale correlations in such a map, along with the above-mentioned small-scale features. We used the latest
available full sky maps, i.e. those of the Wilkinson Microwave Anisotropy Probe (WMAP) of W, Q, V-bands, and the foreground cleaning procedure elaborated by Tegmark et al. (2003). The power spectra obtained for them have common structures that are, however, absent in the simulated maps based on the CMB temperature power spectrum. This is the first attempt, and more detailed analysis of the K-parameter's power spectra can be performed when higher resolution CMB maps are available.

\section{Kolmogorov's stochasticity parameter map}

The Kolmogorov map can be constructed by estimating the stochasticity parameter for the CMB temperature dataset sequence. Kolmogorov's stochasticity parameter \cite{Kolm,Arnold} is defined for the sequence $\{X_1,X_2,\dots,X_n\}$ in increasing order. The cumulative distribution function is
$F(x) = P\{X\le x\}\ $, and the empirical distribution function is defined as

\begin{equation}
F_n(x)=
\begin{cases}
0\ , & x<X_1\ ;\\
k/n\ , & X_k\le x<X_{k+1},\ \ k=1,2,\dots,n-1\ ;\\
1\ , & X_n\le x\ .
\end{cases}
\end{equation}
The stochasticity parameter is 
\begin{equation}\label{KSP}
\lambda_n=\sqrt{n}\ \sup_x|F_n(x)-F(x)|\ .
\end{equation}
The universality of this definition stems from how for any continuous $F$, the convergence 
$
\lim_{n\to\infty}P\{\lambda_n\le\lambda\}=\Phi(\lambda)\ ,
$
where 
\begin{equation}
\Phi(\lambda)=\sum_{k=-\infty}^{+\infty}\ (-1)^k\ e^{-2k^2\lambda^2}\ ,\ \Phi(0)=0,\, \  \lambda>0\ ,\label{Phi}
\end{equation}
is uniform, and $\Phi$ is independent on $F$ \cite{Kolm}.

More specifically, to obtain the degree of randomness for a given sequence, one must compute the Kolmogorov stochasticity parameter $\lambda_n$, and then the estimated Kolmogorov's distribution $\Phi$ will provide information on the degree of randomness in the sequence for the $\lambda_n$ within the interval of their probable values, i.e. approximately, 0.4-1.8 \cite{Arnold_ICTP}. The mean value of $\lambda_n$
given by Kolmogorov distribution is
\begin{equation}
\lambda_{mean}=\int{\lambda\Phi'(\lambda)d\lambda}\approx 0.875029.
\end{equation}
The behavior of $\lambda_n$ and $\Phi$ for a set of sequences that model the CMB as composition of signals, is studied in Ghahramanyan et al. (2009).

The Kolmogorov map obtained based on this definition exhibits, as mentioned above, that structures are linked not only to those noticed by other descriptors but also to those indicating voids \cite{GK2009}.    

\section{Power spectrum}

Once the Kolmogorov statistic $\Phi$ is represented on a map, then one can define
a correlation function on a sphere in spherical coordinates, as for the temperature
data,
\begin{equation}
C(\theta )=<\Phi (\vec{n}_{1})\Phi (\vec{n}_{2})>,\,\,\,\vec{n}_{2}\vec{n}%
_{2}=\cos \theta,
\end{equation}
and expand $\Phi$ via spherical harmonics, 
\begin{equation}
\Phi (\theta ,\varphi )=\sum_{l,m}a_{lm}Y_{lm}(\theta ,\varphi ),
\end{equation}%
where the coefficients $a_{lm}$, as usual, are found from 
\begin{equation}
a_{lm}=\int \Phi (\theta ,\varphi )Y^{\ast}_{lm}(\theta ,\varphi )\sin \theta
d\theta d\varphi .
\end{equation}%
Then  
\begin{equation}
C(\theta )=\frac{1}{4\pi }\sum_{l,m}(2l+1)C_{l}P_{l}(\cos \theta )
\end{equation}%
and 
\begin{equation}
C_{l}=<a_{lm}^{\ast}a_{lm}>  \label{Clr}
\end{equation}%
or 
\begin{equation}
C_{l}=\frac{1}{2l+1}\sum_{m=-l}^{l}|a_{lm}|^{2}  \label{Cl}.
\end{equation}%
However, for our purposes, i.e. when the $\Phi$ is averaged within certain numbers of pixels with noise, the cross-power spectra $\widetilde{C}_{l}^{ij}$ of various bands are more efficient than those of autocorrelations; i.e., then one may get more cleaner power spectrum for correlations, we study the cross power spectra for $\Phi$ 
\begin{equation}
\widetilde{C}_{l}^{ij}=\frac{1}{2l+1}\sum_{m=-l}^{l}a_{lm}^{i}a_{lm}^{j\ast }
\label{Ccross}
\end{equation}%
where $i\neq j, i,j=1,...,8$  for Q1, Q2, V1,V2, and W1-W4 bands. 

In our analysis we used the eight of WMAP's maps, of W, V, Q-bands,  in the usual HEALPix format \cite{HP},
of the resolution parameter $n_s=512$, of a total number of pixels $%
N_{pix}=12n_s^2=3145728$. For each $n_s=512$ temperature map,
we constructed Kolmogorov's stochasticity parameter map for $%
n_s=32$, $N_{pix}=12288$, since for the Kolmogorov map one needs about 100
temperature pixels. To obtain the $n_s=32$ K-map from the 
$n_s=512$ CMB map, each $\Phi$ pixel is calculated from 64 temperature pixels.

Then, for the HEALPix map of given $n_{side}$, the maximum $l$ in the obtained power spectrum will be 
\begin{equation}
l_{max}=\sqrt{3\pi}n_{s}.  \label{lm}
\end{equation}
This corresponds to $l_{max}=96$ for $n_{s}=32$ map and $l_{max}=1536$ for $%
n_{s}=512$. 
The procedure for getting $\Phi$ cross-power spectra included: 

1. calculation of  $i$-th $a_{lm}$ for each K-map, 

2. obtaining of all possible combinations of cross-power spectra,

3. estimation of the mean and the error bars for the set of spectra:

\begin{figure}[ht]
\begin{center}
\centerline{\epsfig{file=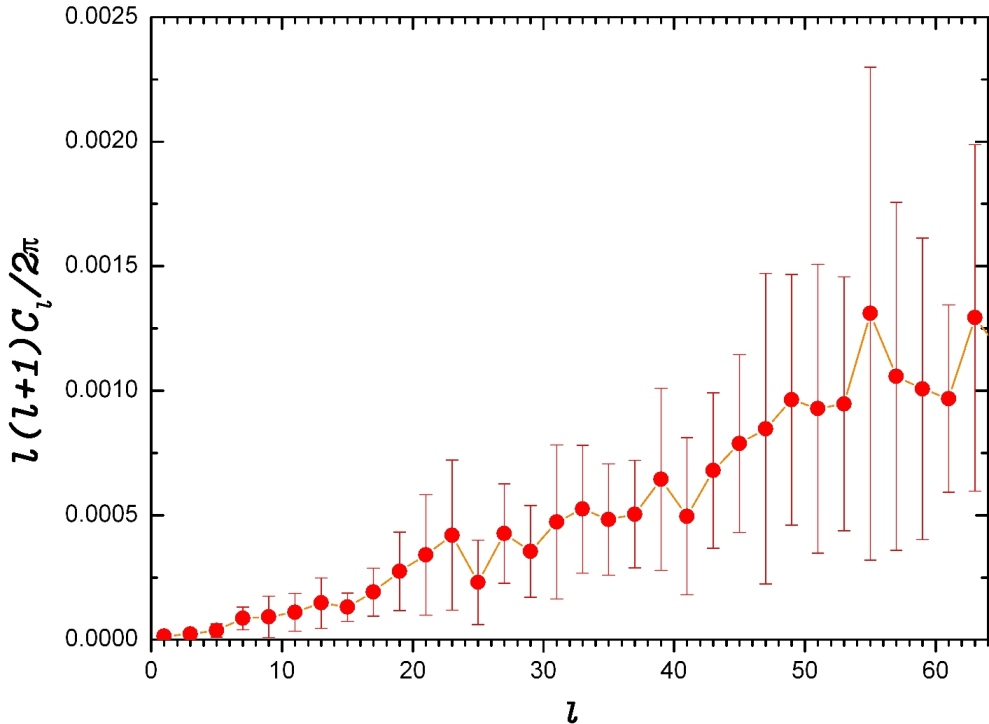,width=0.5\textwidth}} \vspace*{8pt}
\end{center}
\caption{The mean for 15 cross-power
spectra for Kolmogorov CMB maps for WMAP's 6 frequency bands, V1,V2,W1,W2,W3,W4.}
\label{15mean}
\end{figure}

\begin{eqnarray}
\widetilde{C}_{l(mean)} &=&<\widetilde{C}_{l}^{ij}>,\,\,i\neq j,\,0\leq l\leq 96;  \notag \\
\epsilon &=&\sqrt{<(\widetilde{C}_{l(mean)}-\widetilde{C}_{l}^{ij})^{2}>}.  \label{cross}
\end{eqnarray}
Note that $a_{lm}$-s are complex variables, making the correlation function complex as well.
However, since the noise differs from map to map, the resulting complex part
due to noise is vanishing at cross correlations. 
The calculations were performed for $a_{lm}$ without a Galactic disk region within $\pm
20^{\circ}$, for 6 and 8 K-maps, and we get 15 and 28
cross-power spectra, respectively, once their mean and error bars were obtained. The
results are shown in Fig. 1. We see that, for the 28 cross-power spectra, the mean 
is the same as for 15, but the estimated errors are bigger because of using the noisier Q1, Q2 maps.

The mean power spectrum is similar to the CMB pseudo-power spectrum
discussed in Hinshaw et al. (2003), so one may think to use the Peebles weighting
method \cite{Peebles73,Hivon2002} to find the power spectrum with the Galactic disk. However, this causes
two types of difficulties. First, we do not have enough pixels
($n_s=32, N_{pix}=12288$) to calculate the $a_{lm}$ up to $l=250$, which is needed
for calculating the precise weighting. Second, even if we keep the Galactic
disk region where we have approximately $\Phi=1$, it differs very little from the situation
if we a priori adopt $\Phi=1$. The reasonable solution seems not to use
the Galactic region at all and to construct the power spectrum only for odd $l$, which
are not affected by the Galactic disk cut.

\section{Foreground cleaned $\Phi$ map}

We then obtained the power spectrum of $\Phi$ using the foreground cleaning method
developed for CMB maps by Tegmark et al. (2003) and the linear combination method of (Saha et al. 2006, 2008). 
This is based on the use of a linear combination of different maps with weighting of $w_l^i$, not only depending
on $i$-th map but also on the multipole $l$. 
\begin{figure}[ht]
\begin{center}
\centerline{\epsfig{file=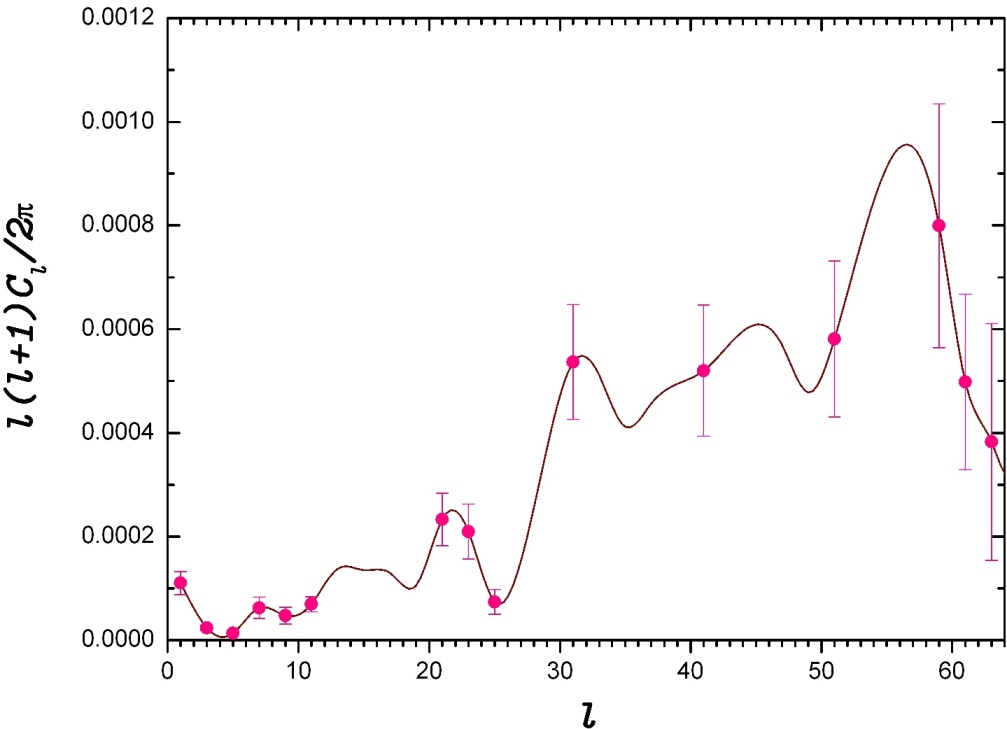,width=0.5\textwidth}} \vspace*{8pt}
\end{center}
\par
\caption{The power spectrum for foreground cleaned Kolmogorov maps.}
\label{ps-final}
\end{figure}

For constructing of a cleaned $\Phi$ map, we first calculated all $a_{lm}$ and then
calculated the cleaned $a_{lm}$ using the relation 
\begin{equation}
a_{lm}^{(clean)}=\sum_{i=1}^{8}w_{l}^{i}a_{lm}^{i}
\end{equation}%
where $w_{l}^{i}$ is 
\begin{eqnarray}
w_{l}^{i} &=&\frac{\sum_{k=1}^{8}e_{k}(C_{l}^{-1})^{ki}}{%
\sum_{i,k=1}^{8}e_{k}(C_{l}^{-1})^{ki}e^{i}},  \notag \\
\sum_{i=1}^{8}{w_{l}^{i}} &=&1.  \label{wl}
\end{eqnarray}%
Here $C_{l}$ is an $8\times 8$ dimensional matrix constructed by all possible
auto and cross-power spectra from all maps (see eq.(\ref{Ccross})), so that $%
C_{l}^{-1}$ refers to an inverse matrix, $e^{i}$ and $e_{i}$ are 
8-dimensional unit vector and its transponated vector, respectively. For the covariance of
this representation see (Tegmark et al. 2003). We get all power spectra
in Eq.\ref{wl} smoothed by $\Delta {l}=10$ interval to avoid the
singular $C_{l}$ matrix. For example, we get triplets of different maps from different
bands Q1,V1,W1 for $w_{l}^{i}$. 

We thus get 16 different triplets. For any triplet, a linearly superposed $a_{lm}$-s is
constructed. The last step is to find all possible cross-power spectra
from those linearly superposed ones, whose initial map components are different. For
example, $(Q1+V1+W1)\otimes (Q2+V2+W2)$ complies to this restriction, but $%
(Q1+V1+W1)\otimes (Q2+V2+W1)$ does not, so only 3 maps of eq.(\ref{wl}) were used. In this way we obtain three cross-power spectra from triplets. The mean power spectra from these
cross-power spectra is shown in Fig. 2.

\section{Simulations}

We repeated the estimations  described above for simulated maps. 
Four different types of simulations were constructed from: 

a.  the real maps'  $a_{lm}$-s (T maps),

b. real maps with added Gaussian noise
of the same parameters as the noise in WMAP CMB maps (T+N maps),

c. Gaussian maps of the distribution
parameters $T,\sigma$ from WMAP W band real map (G maps), and

d. Gaussian  maps with added Gaussian noise, both
from the parameters $T,\sigma$ of WMAP W band map
(G+N maps). 
\begin{figure}[ht]
\begin{center}
\centerline{\epsfig{file=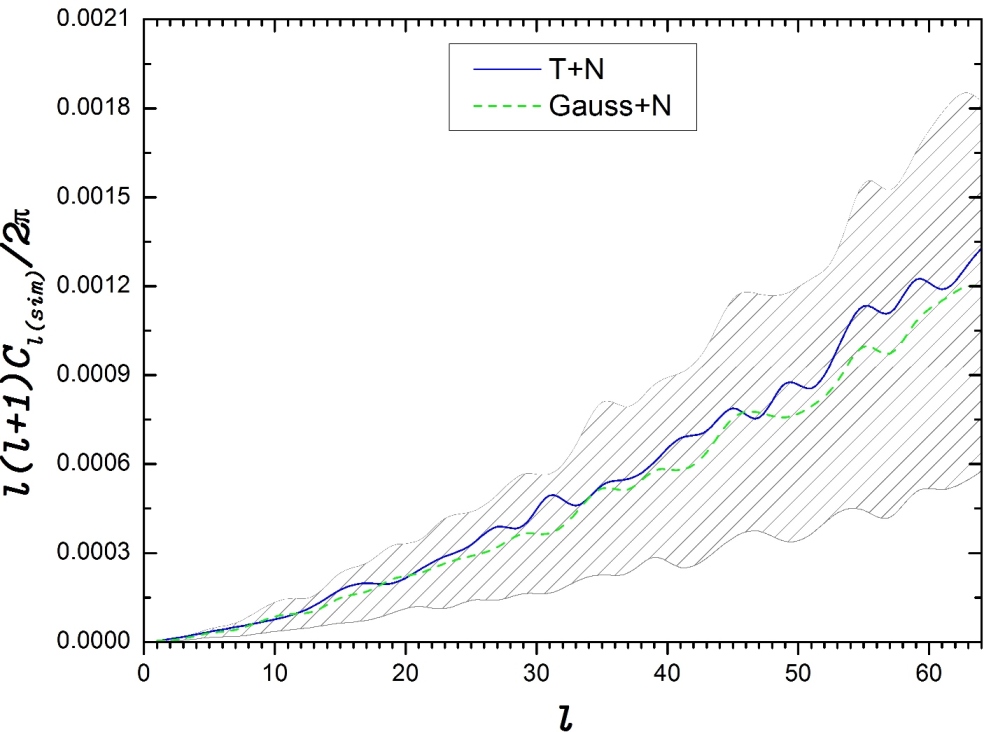,width=0.5\textwidth}} \vspace*{8pt}
\end{center}
\par
\caption{The Kolmogorov power spectra for simulated CMB temperature (T+N) and Gaussian (G+N) maps with superimposed WMAP's noise, averaged over 190 cross-power spectra each; the smoothed error bars are shown.}
\label{SimT}
\end{figure}

For each group of $n_s=32$ $\Phi$ simulated map we obtain the mean cross power spectra as described above. For 20
different maps one has 190 cross power spectra. Similarly, 190 cross-power spectra were computed for the Gaussian maps generated with the WMAP's $\sigma$ and mean temperature and with superimposed noise of WMAP. 

Although the number of the cross spectra for simulations is more than those we used for calculating the power spectra for real K-maps, neither of the resulted spectra shows the features found for real maps with 0.6 and 2.7-$\sigma$ level for W and foreground cleaned maps, respectively, as shown in Fig. 3. Even more important than the $\sigma$-level, however, seems that the features only appear at cross and not at auto correlations, thus indicating that they do not come from the noise in the maps. The principal limitation in the above analysis is the angular resolution, since $\Phi$ reflects the statistical properties of the signal, the efficiency of the method will increase with higher resolution data.  

\section{Conclusion}

We have obtained the first power spectrum of Kolmogorov stochasticity parameter map of CMB temperature data.
The WMAP W,Q,V-band datasets were used to compute the Kolmogorov's CMB maps. The foreground cleaning method of Tegmark  et al. (2003) was also applied while computing the $\Phi$ maps. The mean for the set of cross-correlated maps
was computed. They show features, particularly at around $l=25$, that are absent in the maps simulated either for the WMAP's temperature power spectrum parameters or in the Gaussian maps with superimposed noise, i.e. additional effects  to those usually included in the simulated models. 

Although the accuracy of the present analysis is limited by the WMAP's angular resolution and signal-to-noise ratio, it shows the principal possibility of obtaining such crucial information from CMB, and even the already obtained behaviors can affect the development of scenarios for the void correlations at the large-scale structure formation.

Higher angular resolution maps expected soon at Planck and other experiments will enable the finer structure analysis of structures in the power spectra of Kolmogorov CMB maps.

\end{document}